# Momentum-dependent snapshots of a melting charge density wave

J.C. Petersen[1,2], S. Kaiser[2], N. Dean[1], A. Simoncig[2], H.Y. Liu[2], A.L. Cavalieri[2], C. Cacho[3], I.C.E. Turcu[3], E. Springate[3], F. Frassetto[4], L. Poletto[4], S.S. Dhesi[5], H. Berger[6], and A. Cavalleri[1,2,*]

[1]*Department of Physics, University of Oxford, Clarendon Laboratory, UK*

[2]*Max Planck Department for Structural Dynamics, CFEL, Hamburg, Germany*

[3]*Central Laser Facility, STFC Rutherford Appleton Laboratory, Didcot, UK*

[4]*CNR-Institute for Photonics and Nanotechnologies, Padova, Italy*

[5]*Diamond Light Source, Harwell Science and Innovation Campus, Chilton, Didcot, UK*

[6]*École Polytechnique Fédérale de Lausanne, Lausanne, Switzerland*

**Charge density waves[1] (CDWs) underpin the electronic properties of many complex materials[2,3,4,5]. Near-equilibrium CDW order is linearly coupled to a periodic, atomic-structural distortion, and the dynamics is understood in terms of amplitude and phase modes. However, at the shortest timescales lattice and charge order may become de-coupled, highlighting the electronic nature of this many-body broken symmetry ground state. Using time and angle resolved photoemission spectroscopy with sub-30-fs XUV pulses, we have mapped the time- and momentum-dependent electronic structure in photo-stimulated 1$T$-TaS$_2$, a prototypical two-dimensional charge density wave compound. We find that CDW order, observed as a splitting of the uppermost electronic bands at the Brillouin zone boundary, melts well before relaxation of the underlying structural distortion. Decoupled charge and lattice modulations challenge the view of Fermi Surface nesting as a driving force for charge density wave formation in 1$T$-TaS$_2$.**

At room temperature, $1T$-$TaS_2$ is a metal, exhibiting a nearly commensurate charge density wave (CDW) [6,7]. Below 180 K the CDW becomes fully commensurate and locks into the lattice[8,9,10], concomitant with a long-range periodic distortion of the crystal structure (Fig. 1 (b)). This lattice distortion reduces the size of the Brillouin zone and leads to back-folding of the Ta $5d$ manifold, splitting off several Umklapp or 'shadow' bands. Gaps appear at the new band crossings, and only a narrow band is left at the Fermi level[11,12]. This occurrence of at least three different phenomena at once – lattice modulation, charge ordering, and Mott localization – opens many questions on the exact relationship between them, which must be answered in order to understand the dynamics in this compound.

Excitation with ultrashort pulses of light has been shown to promptly close the gap at the zone centre[13], highlighting the importance of Mott physics in the insulating ground state. Photo-excitation also couples to the amplitude mode of the CDW[14], a 2.4-THz-frequency oscillation connected to breathing of both the star-shaped atomic clusters and of the charge density. Time-resolved electron diffraction experiments[15,16] have shown that the long-range atomic-structural order is not completely relaxed, whilst time resolved core level photoemission measurements suggest that the onsite charge density is already perturbed within 100 fs[17], implying that charge and lattice may be decoupled at the earliest timescales.

Optical experiments with sub-vibrational time resolution[18,19,20] have shown that the nature of a complex insulator is reflected directly in the speed at which its electronic structure rearranges after a prompt change in filling, but optical spectroscopy alone lacks the momentum sensitivity required to fully characterise the transient electronic structure. To address this issue, it is necessary to measure the electronic structure at the boundary of the Brillouin zone, observing the time dependent Umklapp-band splitting as a direct indication of charge order.

Experimentally, measurements of this kind have not been possible to date. Pioneering time resolved photo-emission measurements have typically used relatively long pulses (100 fs or longer) and only near UV radiation (< 7 eV), making it difficult to access sub-vibrational timescales and mapping electronic structures over only limited portions of momentum space[21,22,23]. In the present study, we perform time and angle resolved photo-emission spectroscopy using XUV pulses from high-order laser harmonics, with a temporal resolution of 30 fs (see Methods section).

Figure 1 (a) shows static photoemission intensity maps of the band structure near the Fermi level, $E_f$, measured for reference with an ultraviolet helium lamp below and above the thermal phase transition at 180 K. In the low-temperature phase, reduced photoemission intensity at the Fermi energy indicates the opening of the Mott gap. The band immediately below $E_f$ is the lower Hubbard band (LHB). A gap between sub-bands is evident at the reduced Brillouin zone boundary (the M point). High-momentum states are folded back into the first (reduced) zone, so the split-off bands reappear as Umklapp bands at $k$=0 (the Γ point). We call the uppermost of these U1. Each of these features is reproduced in the static photoemission intensity map obtained with our laser-based XUV source in Fig. 1 (c). The quantitative differences between the two static spectra arise from the differences between the sources: the He lamp is not polarized whilst the high-order harmonic source is *p*-polarized, and the photon energies are slightly different.

To measure the changes in electronic properties after photostimulation, a 30-fs pulse of infrared light from an 800-nm-wavelength laser excites the surface of a sample held at 20 K in a UHV chamber. A time delayed XUV pulse with photon energy of 20.4 eV, produced by high-harmonic generation, generates photoelectrons whose energy and momentum are determined by a hemispherical analyser. By varying the delay between the infrared and XUV pulses, we are able to follow the electronic structure as a function of time during and after the photoinduced insulator-to-metal transition.

Figure 2 shows the energy distribution of intensity at the Γ and M points as a function of time. At Γ, a rapid loss of intensity in the LHB confirms that the Mott gap collapses, accompanied by filling in of states at and above the Fermi level. This is followed by recovery of the original peak, followed by a rigid oscillation of the band edge at the frequency of the CDW amplitude mode. These observations are in accord with earlier results[13,21], where 6-eV photon energy pulses were used to study states near $k$=0. At the M point, there is a similar rapid loss of overall intensity, transfer across $E_f$, and recovery. In addition, the splitting between U1 and LHB is destroyed, as clearly visible in the individual energy spectra shown in Fig. 2 (b). These prompt changes are followed by the rapid recovery of both gaps. The entire LHB/U1 complex then oscillates at the frequency of the CDW amplitude mode[13], as does the splitting between sub-bands.

Next, we set out to investigate the detailed dynamics and $k$-dependence of the changes in the LHB and U1 bands during the formation of the photometallic phase. The destruction of both gaps is clearly evident on sub-vibrational time scales. Figure 3 shows maps of the photoemission intensity at various times during the ultrafast phase transition. Again, intensity is lost at the peak of the LHB and transferred across the Fermi level as a result of the melting of the Mott state. Simultaneously, the gap between sub-bands disappears at the M point, indicating that the CDW order has melted as well. However, not all changes to the electronic structure proceed so promptly. Figure 4 (a) shows a more detailed examination of the time-dependence at Γ. At the Fermi level, where the increase in intensity corresponds to the collapse of the Mott gap, the change is expected to proceed on the time scale of hopping[20]. The transition here is effectively prompt, occurring on a timescale faster than the 30-fs resolution of the experiment. The same process drives the initial changes at the peak of the LHB, which also rearranges promptly.

In contrast to the prompt loss of the U1-LHB splitting, at the lower energies of the U1 band at Γ the response is slower, well described by a transient response with a timescale of a half-cycle of the amplitude mode. This is consistent with a reduction in spectral weight in the Umklapp band brought about by incipient relaxation of the lattice distortion, which proceeds on structural rather than electronic time scales[16].

To illustrate this idea further, Fig. 4 (b) shows the changes in intensity across the entire range of momenta. The first panel shows the relative changes from 50 fs before photoexcitation to 50 fs after, while the second panel shows the changes between 50 and 200 fs after photoexcitation. The changes on short timescales reiterate that as a consequence of photo-induced melting of the Mott state, spectral intensity is lost at lower energies – especially at the peak of the lower Hubbard band – and transferred across the Fermi level. This process is largely independent of momentum, indicating that the effects of electron-electron interaction pervade the Brillouin zone.

In contrast with the prompt response of the Mott gap, dispersive electronic structural changes are seen at longer time delays, and two key features are apparent. Intensity is lost in U1 at Γ, with this intensity reappearing at Γ', providing a direct snapshot of the unfolding of the Brillouin zone. Overall, increased intensity is mostly located above the LHB peak, and decreased intensity below it, indicating that the band edge is moving upward as the structure begins to relax.

The fact that these different processes take place on distinctly electronic and structural timescales gives new information about both the static band structure of 2D CDW materials, and the process by which they melt after ultrafast excitation. The prompt loss of the gap between sub-bands reflects the broken symmetry of charge density in the CDW ground state, which appears to be connected to electronic correlations. On the contrary, the intensity of the shadow bands at the zone centre must be a manifestation of the periodic lattice distortion. The ordered sequence of spectral changes also indicates that the CDW melts via prompt destruction of the charge order, to which the lattice then responds on its own timescale by relaxation of the periodic lattice distortion. This unifies previous experimental results which have shown rapid melting of a CDW gap[24] and slow melting of lattice order[16] in different materials, by simultaneously capturing the dynamics of all these features in the same material at the same time.

In summary, we have extended the technique of time-resolved ARPES beyond its previous limits to study the relaxation of a charge density wave in photo-stimulated 1$T$-TaS$_2$, measuring charge order at the edge of the Brillouin zone with sub-vibrational time resolution. Melting of the charge density wave is prompt within the 30 fs resolution of our measurement, as is the loss of the CDW-induced gap between sub-bands. Meanwhile, the Brillouin zone unfolds on the lattice time scale, consistent with the notion that a 20% lattice distortion along the coordinate of the Raman active amplitude mode follows photo-doping[16]. At longer timescales, charge order and lattice distortions lock again, displaying the well-known modulations of spectral weight occurring at the frequency of the amplitude mode. These measurements highlight the importance of electron-electron correlations in localizing the charge on each site, and challenge the prevailing view of CDW order in 1$T$-TaS$_2$ as driven by Fermi surface nesting.

**Methods**

In angle-resolved photoemission spectroscopy (ARPES), photons illuminate the sample surface and stimulate the emission of photoelectrons, which are collected and analysed. In a quasi-two-dimensional material, the angle at which an electron exits the sample surface (along with the electron energy) can be converted directly to the momentum of the state that had been occupied inside the material. Regions in energy-momentum space with high photoemission intensity correspond to occupied states within the material, and therefore reveal the electronic structure[25,26].

The experiments reported here employed the time-resolved laser ARPES setup at the Artemis facility, in the Central Laser Facility of the Rutherford Appleton Laboratory. A 1-kHz pulsed beam of 30-fs pulses from an amplified 800-nm laser is split into two parts. One is used as the pump, that is, to drive the system out of its equilibrium state. The other enters a beamline held under vacuum, and is focussed into a pulsed jet of argon gas in order to generate XUV pulses via the non-linear optical process known as high-order harmonic generation. The XUV passes through a grating monochromator designed to preserve the pulse duration[27], which selects the 13$^{th}$ harmonic (photon energy 20.4 eV). A grazing-incidence toroidal mirror focuses the pulsed XUV onto the sample, which is held on a liquid-helium-cooled sample manipulator in a UHV chamber. Photoelectrons emitted from the sample surface after each XUV pulse are collected and measured by a hemispherical analyser (SPECS PHOIBOS 100) with a multi-channel plate and CCD detector. The overall energy resolution of the system is approximately 150 meV, limited by the line width of the laser XUV source. Improvements to this energy resolution would require the sacrifice of time resolution.

$TaS_2$ samples are cleaved under UHV while cold, and surface quality is inspected by LEED before and after ARPES runs.

**Acknowledgements**

We thank S. Hook and T. Strange for technical support, and C. Froud and W.A. Bryan for assistance with the development of the beamline. This research received funding from the EC Seventh Framework programme via Laserlab Europe.

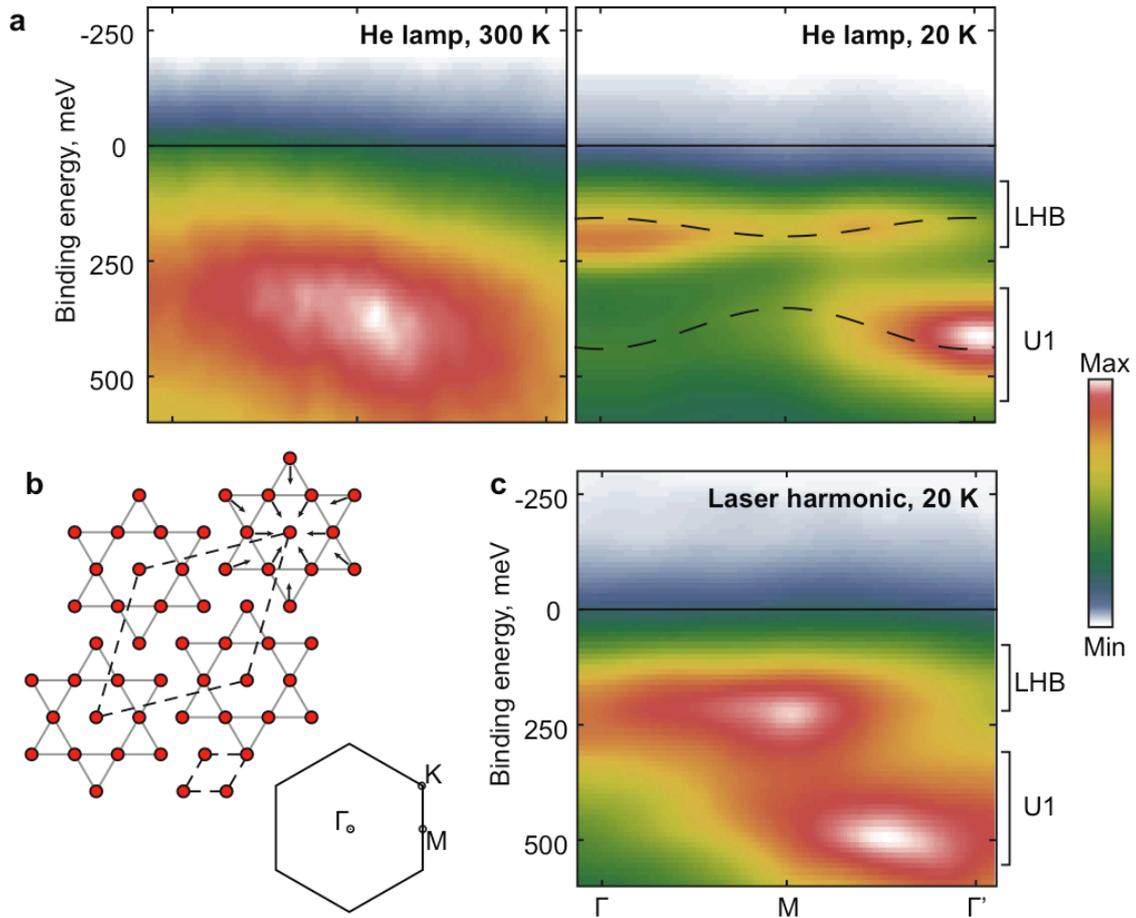

**Figure 1: Structure and ARPES intensity maps of 1$T$-TaS$_2$. a,** Static photoemission intensity maps from centre to centre of the first two Brillouin zones, measured with a helium discharge lamp (21.2 eV) at 20 K and 300 K. A solid line denotes the Fermi level. Dashed lines indicate the approximate band structure in the low-temperature phase, showing the positions of the LHB (upper band) and U1 (lower band). In both panels, momentum labels refer to the Brillouin zone of the low-temperature structure. **b,** Low-temperature lattice distortion in the Ta plane, which accompanies the formation of the commensurate, static CDW. Dashed lines indicate high- and low-temperature unit cells. Grey stars represent clusters of atoms formed by the periodic lattice distortion, illustrated by arrows for one cluster. Also shown is a map of the high-symmetry points of the Brillouin zone. **c,** static intensity map measured with the laser-harmonic XUV source.

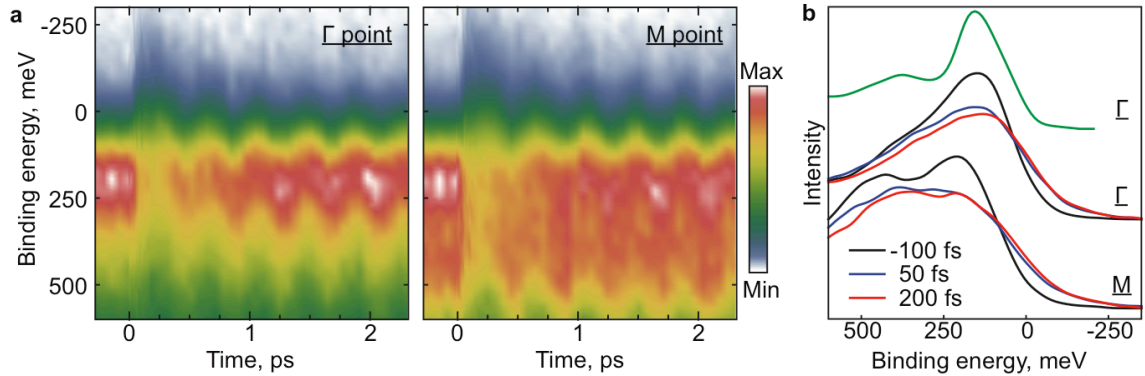

**Figure 2: Collapse of the Mott and CDW gaps and oscillation of the LHB edge. a,** Maps of photoelectron intensity as a function of time, during and after photoexcitation. At both the Γ point and the M point, a prompt overall transfer of intensity across the Fermi level heralds the destruction of the Mott gap. At the M point, the two distinct peaks of the lower Hubbard and uppermost Umklapp bands momentarily collapse into a single feature. A rapid recovery and subsequent rigid oscillation at the amplitude-mode frequency follows at both Γ and M. **b,** Individual energy distribution curves before and after photoexcitation, showing the rapid loss of splitting between the band features at the M point. The top curve (green) is a reference measurement taken with a lamp source, showing the two bands at the zone centre.

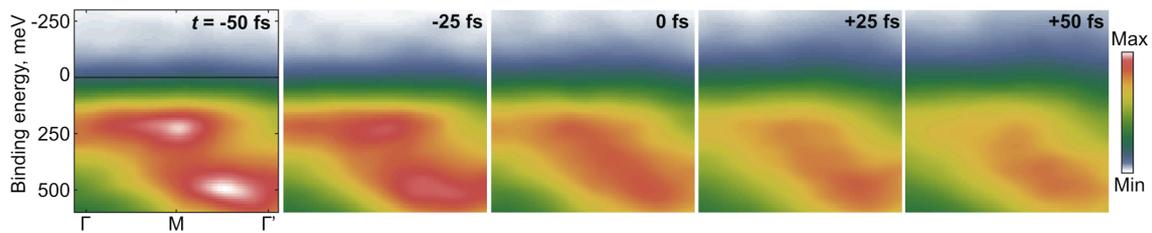

**Figure 3: Snapshots of momentum-dependent photoemission intensity,** obtained with the laser-harmonic XUV source at 20 K. Dashed lines in the first panel denote the equilibrium band positions. As the insulator-metal transition proceeds, transfer of intensity across the Fermi level is evident throughout both Brillouin zones. The gap between bands at the zone boundary collapses on sub-vibrational timescales.

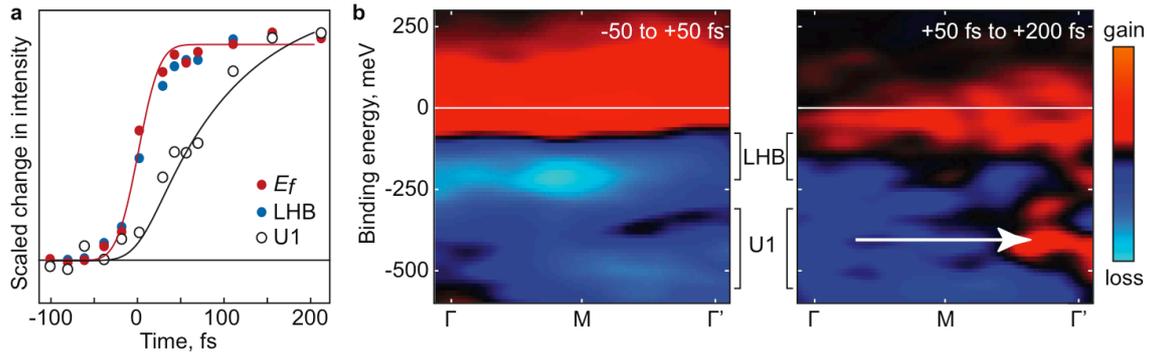

**Figure 4: Momentum dependence of spectral changes on fast and slow timescales. a,** Time evolution at the Γ point, scaled, for three key energies. Changes are prompt in the lower Hubbard band and at $E_f$, but much slower in the Umklapp band (U1). Red and black curves show a step response and a transient response with a lifetime of half the amplitude-mode period, each convolved with the instrumental resolution. **b,** Maps of intensity changes. The first panel shows immediate changes, between -50 and +50 fs. The second panel shows the subsequent change, between +50 and +250 fs. Changes at early times are characterised by a loss of intensity in each band, and a gain in intensity at the Fermi level and above. On the slower timescale, intensity is lost below the LHB and gained above it, as the peak shifts upwards. In U1, shadow-band intensity is lost in the first zone and transferred to the second, as the structure relaxes and the Brillouin zone unfolds (solid arrow).

# References

* to whom correspondence should be addressed: andrea.cavalleri@mpsd.cfel.de and j.petersen1@physics.ox.ac.uk